\documentclass[a4paper,fleqn]{cas-sc}

\usepackage[numbers]{natbib}

\def\tsc#1{\csdef{#1}{\textsc{\lowercase{#1}}\xspace}}
\tsc{WGM}
\tsc{QE}

\begin{document}
\let\WriteBookmarks\relax
\def\floatpagepagefraction{1}
\def\textpagefraction{.001}

\shorttitle{Multi-parameter tests of GR with PCA for LISA}    

\shortauthors{R. Niu, Z.-C. Ma, J.-M. Chen, C. Feng, W. Zhao}  

\title [mode=title]{Multi-parameter Tests of General Relativity Using Bayesian Parameter Estimation with Principal Component Analysis for LISA}  



%

\author[1, 2]{Rui Niu}[orcid=0000-0001-9098-6800]


\ead{nrui@mail.ustc.edu.cn}

%

\affiliation[1]{organization={CAS Key Laboratory for Researches in Galaxies and Cosmology, Department of Astronomy, University of Science and Technology of China, Chinese Academy of Sciences},
                city={Hefei},
                postcode={230026}, 
                state={Anhui},
                country={China}}
\affiliation[2]{organization={School of Astronomy and Space Sciences, University of Science and Technology of China},
                city={Hefei},
                postcode={230026}, 
                state={Anhui},
                country={China}}
\affiliation[3]{organization={North Information Control Research Academy Group Co. Ltd.},
                city={Nanjing},
                postcode={211153}, 
                state={Jiangsu},
                country={China}}

\cortext[1]{Corresponding author}

\author[1, 2]{Zhi-Chu Ma}[orcid=]
\ead{zcma@mail.ustc.edu.cn}

\author[3]{Ji-Ming Chen}[orcid=]
\ead{chenjm94@mail.ustc.edu.cn}

\author[1, 2]{Chang Feng}[orcid=0000-0001-7438-5896]
\cormark[1]
\ead{changfeng@ustc.edu.cn}

\author[1, 2,]{Wen Zhao}[orcid=0000-0002-1330-2329]
\cormark[1]
\ead{wzhao7@ustc.edu.cn}



\begin{abstract}
In the near future, space-borne gravitational wave (GW) detector LISA can open the window of low-frequency band of GW and provide new tools to test gravity theories.
In this work, we consider multi-parameter tests of GW generation and propagation where the deformation coefficients are varied simultaneously in parameter estimation and the principal component analysis (PCA) method are used to transform posterior samples into new bases for extracting the most informative components. 
{The dominant components can be more sensitive to potential departures from general relativity (GR). }
We extend previous works by employing Bayesian parameter estimation and performing both tests with injections of GR and injections of subtle GR-violated signals. 
We also apply multi-parameter tests with PCA in the phenomenological test of GW propagation. 
This work complements previous works and further demonstrates the enhancement provided by the PCA method. 
Considering a supermassive black hole binary system as the GW source, 
{we show that subtle departures will be more obvious in posteriors of PCA parameters.}
The departures less than $1\sigma$ in original parameters can yield significant departures in first 5 dominant PCA parameters.
\end{abstract}




\maketitle

\section{Introduction} \label{sec_intro}

General Relativity (GR) is regarded as the most successful gravity theory due to its elegant mathematical formulation and agreement with experiments at extremely high precision \citep{Berti2015,Will2014,Hoyle2001,Adelberger2001,Jain2010,Koyama2016,Stairs2003,Manchester2015,Wex2014,Kramer2017}.
Whereas, difficulties of singularity and quantization problems \citep{DeWitt1967,Kiefer2007}, as well as  puzzles of dark matter and dark energy \citep{Cline2013,Sahni2004,Debono2016,Solomon2019} hint that GR may be not complete to describe all gravitational phenomena, which motivates people to construct alternatives theoretically and search anomalies experimentally \citep{Sathyaprakash2019, Clifton2012, Uzan2010}. In previous, extensive tests of GR have be performed from the laboratory scale \citep{Sabulsky2019,Hoyle2001,Adelberger2001}, to the solar system scale \citep{Will2018,Will2014}, and to the cosmological scale \citep{Jain2010,Koyama2016,Clifton2012}. 
In recent years, the successful detection of gravitational waves (GWs) provide a new window to observe the universe and offer an unique tool to test gravity theories \citep{Collaboration2021g, Collaboration2020, Abbott2019b, Abbott2019c, Abbott2016a}. 

Since the first detection of GW150914 \citep{Abbott2016}, more than 90 confident GW events have been captured by the LIGO, Virgo, KAGRA (LVK) Collaboration \citep{Abbott2019,Abbott2020,Collaboration2021e, Collaboration2021f}. Extensive tests of GR with GW are flourishing based on observed data from current ground detectors, such as \citep{Perkins2021,Wang2021d,Haegel2023,Gong2022,Wu2022}.
In the near future, the space-borne GW observatories the Laser Interferometer Space Antenna (LISA), Taiji, and Tianqin are scheduled to be launched in the early 2030s \citep{Ruan2020, Luo2016, AmaroSeoane2017a}. Space-borne detectors can open the window of low-frequency GW around milli-Hertz range which encompasses abundant GW sources like supermassive black hole binaries (SMBHBs) \citep{Klein2016}, extreme mass ratio inspirals (EMRIs) \citep{Babak2017a}, etc. These sources can offer invaluable information for exploring the nature of gravity \citep{Barausse2020a,Arun2022}. 
Forecast researches about testing GR with space-borne detectors are critical and pressing before its launch.
In this work, we consider LISA as the example to discuss phenomenological parameterized tests of GW generation and propagation.

Tests of gravity theories with GW are generally approached from three perspectives, the generation, propagation, and polarization of GWs \citep{Gair2013, Will2014}. 
In this paper, we focus on the generation and propagation of GWs. 
In alternative theories of gravity, the energy and angular momentum of binaries, as well as flux of them are usually different with GR \citep{Will2018}, which can result in modifications to the binary motion and yield deformations in the waveform. 
For example, the dipole radiation is absent in GR but very common in plethora alternative theories \citep{Yagi2016}.
In GR, GWs propagate non-dispersively with the speed of light, whereas in alternative theories this feature can be violated in various ways. Such as in Lorentz violating theories, different frequency components of GWs can propagate with different speed, which causes dispersion and distorts signals observed by detectors \citep{Zhao2020a,Mirshekari2012,Will1998}.

Proposals to test gravity theories in various literature may be classified into two types, performing tests and constraining model parameters in particular alternative theories \citep{Perkins2021,Niu2020,Wang2023a} or testing in model-independent ways \citep{Krishnendu2021a} which aims at searching any indications for deviations from predictions of GR. 
Since the GW waveform in particular theories is usually difficult to compute, and considering the abundance of alternative theories, theory-agnostic tests might be a more efficient way\citep{Krishnendu2021a}. 
In this work, we consider two phenomenological parameterized theory-agnostic tests of GW generation and propagation which have been routinely performed on observed data by LVK \citep{Abbott2019b,Collaboration2020,Collaboration2021g}.
But different with the tests implemented by LVK where phenomenological non-GR coefficients are varied separately in parameter estimation, we consider them simultaneously and employ principal component analysis (PCA) to transform posteriors of orignal modification coefficients into a set of new bases. The dominant components in the obtained PCA parameters are considered to be more sensitive to potential deviations from GR \citep{Datta2020,Pai2012,Saleem2021,Datta2022,Datta2023}.
Tests with varying all deformation coefficients simultaneously are called multi-parameter tests \citep{Saleem2021}.

{Although previous works \cite{Perkins2022,Sampson2013,Meidam2018} have demonstrated that single-parameter tests where only one deformation parameters is allowed to vary can effectively detect deviations from GR, multi-parameter tests are still useful.
On the one hand, in single parameter tests, posteriors only contains information from one deformation parameter. While, in multi-parameter tests, posteriors can collect information from all PN order simultaneously.
Especially considering that space-borne detectors can detect SMBHBs with very high SNR. The contributions from high order terms may become more important.
Collecting more information may help people to identify more subtle deviations. 
On the other hand, all deformation coefficients are possible to departure from GR and need to be considered as free parameters in estimation.
Single parameter tests are equivalent to use a prior where only the sampled parameter is uniformly distributed while other deformation parameters have a delta function prior. In most general case, we may prefer the most agnostic priors where all deformation parameters are uniformly distributed in the possible parameter space.}

Using the PCA method in the multi-parameter test of Post-Newtonian coefficients with GW was first introduced in the previous work \citep{Pai2012}. 
A new set of coefficients can be constructed by a linear combination of original phase deformation coefficients. The new parameters can circumvent correlations among originals, and can be estimated with improved accuracy. 
In the paper \citep{Saleem2021}, authors performed this method with selected GW events detected in first and second observing runs (O1 and O2) of LVK, and simulated non-GR signals, which shows new coefficients constructed by PCA are more sensitive to potential deviations from GR in the multi-parameter test, subtle anomalies can become significant in the posteriors of the new constructed coefficients. 
The works \citep{Datta2022,Datta2023} using Fisher method forecast performance of multi-parameter tests with PCA for future space-borne detector LISA and next generation ground-based detectors Cosmic Explorer (CE) and Einstein Telescope (ET).
In this work, we extent previous work \citep{Datta2023} by employing Bayesian parameter estimation.
Furthermore, We also apply this method in the test of GW propagation.

Previous work \citep{Datta2023} consider the multi-parameter test with PCA using Fisher matrix in which the posteriors are assumed to be Gaussian. However, as pointed out in \citep{Marsat2021}, the multi-modal is one of important features of posteriors for LISA. Therefore, in order to demonstrate performance of multi-parameter tests with PCA more accurately and convincingly, we employ Bayesian inference to estimate posteriors in this work. 
There are various difference between space-borne detectors and ground-based detectors in parameter estimation \citep{Marsat2018}. 
The response function of space-borne detectors have to account for the motion of the detector and the finite dimension of arm-length. 
Since the duration of GW signals for space-borne detectors can be months or years, the motion of detectors can not be ignored. 
In the mHz band, the wavelength of GW could be comparable or less than the arm-length of detectors. Unlike the situation of ground-based detectors where a detector can be viewed as a point, one have to consider the integral of metric perturbation alone geodesic of photons for space-borne detectors.
For unequal-arm interferometers, which is necessarily the case of space-borne detectors, the laser noise experience different delays for two arms, and can not be canceled out at the photon detector. The technique called time delay interferometer (TDI) has to be used to suppress laser noise.

These factors together with longer signal duration make likelihood evaluation for space-borne detectors is much more time consuming.
Various methods have be adopted in previous works to alleviate the computational burden of Bayesian parameter estimation with space-borne detectors. 
In the work \citep{Cornish2020}, the authors employed the heterodyned likelihood \citep{Cornish2010} to speed up the likelihood evaluation. By introducing a reference waveform, the likelihood can be separated as a part of slow-varying function of frequency which can be computed by a coarse interpolation, an integral of a rapidly oscillating function within fast damped envelope which can be truncated at small fraction of its full extent, and a part which can be computed ahead of Markov chain Monte Carlo (MCMC) sampling.
The work \citep{Katz2020} utilized contemporary powerful computer hardware to evaluate full likelihood in "brute-force". The authors implement the codes for waveform model and detector response function on GPU which can generate GW waveform and compute likelihood on a realistic Fourier bin width for the LISA mission in parallel and significantly reduce required computational time.
In this work, we follow the method used in \citep{Marsat2021} which considers the noise-free likelihood and using interpolation to accelerate the likelihood computation.
Furthermore, data from space-borne detectors are expected to be signal-dominant. Heavily overlapped signals bring new challenges to extract science information from data. Global fitting \citep{Karnesis2023,Littenberg2020,Littenberg2023} or hierarchical fitting \citep{Lu2022,Zhang2021a} have to be used to get properties of binaries. 
Issues about overlapping are still under active investigation\citep{Karnesis2023},
therefore in this work we consider an ideal situation where the GW event has been searched out and the segment of data only contains one GW events.

In this work, we perform multi-parameter tests of GR with PCA method for LISA. We extend previous work \citep{Datta2023} by using Bayesian parameter estimation and considering the test of GW propagation.
The remainder of this paper is organized as follows.
In section \ref{sec_method}, we present a brief overview of phenomenological parameterized tests of GW generation and propagation, as well as Bayesian parameter estimation for LISA and the PCA method used to transform posteriors of deformation coefficients into a set of new bases. We consider two situations including tests with injections of GR and injections of subtle deviations from GR to elaborate the advantage of the PCA method in multi-parameter tests. Results are shown and discussed in Section \ref{sec_res}. Finally, we summarize this paper in Section \ref{sec_summary}.

\section{Methodology} \label{sec_method}
In this work, we consider the phenomenological parameterized tests of GW generation and propagation which have been extensively performed with current observed GW data by LVK
\citep{Collaboration2021g, 
Abbott2016a, Abbott2019c, Abbott2019b, Collaboration2020}.
In these tests, several phenomenological coefficients, Eq. \ref{tgr_para_gen} and \ref{phase_deformation_prop}, are introduced to capture any potential deviations from GR. Due to the correlations among parameters, allowing all these coefficients to vary simultaneously in the parameter estimation will yield less informative posterior. In the tests performed by LVK, only one coefficient is allowed to vary at a time.

Previous works \citep{Pai2012} pointed out that such correlations can be reduced by transforming the deformation coefficients into a set of new orthogonal bases obtained from PCA, which allow people to perform multi-parameter tests, i.e. estimating all deformation coefficients simultaneously. 
The effectiveness of multi-parameter tests with PCA has been shown in \citep{Saleem2021, Shoom2023} in which tests of GW generation using GW data detected by current ground based detectors are demonstrated. Using the Fisher method, previous works \citep{Gupta2020,Datta2022,Datta2023,Datta2020} present the forecasts of multi-parameter tests considering future space based detectors, next generation ground based detectors, and multi-band observations through synergies of both.

In this paper, we extend previous works by using Bayesian parameter estimation to further illustrate that the PCA have more sensitivity to potential deviations in multi-parameters tests, and including the phenomenological parameterized test of GW propagation.

\subsection{Multi-parameter tests of GW generation and propagation} \label{subsec_tgr}
In the early inspiral period where orbital velocities of binaries are sufficiently low, motion of compact binaries can be well described by post-Newtonian (PN) formalism in which GW waveform is given by the form of expansion in terms of $v=(\pi M f )^{1/3}$ where $M$ is the total mass of the binary\citep{Blanchet2014}. 
Since potential deviations in GW phase can be accumulated as orbital evolution, GW observations are more sensitive to phase than amplitude in general.
Tests of GW generation usually focus on GW phase.
The 3.5 PN phase of GW waveform \citep{Khan2016} has the form of
\begin{equation}
\Phi_{\mathrm{PN}}(f)= 2 \pi f t_{\mathrm{c}}-\varphi_{\mathrm{c}}-\frac{\pi}{4} +\frac{3}{128 \eta}v^{-5} \sum_{i=0}^{7}\left[\varphi_{i}+\varphi_{i l} \log v\right]v^{i}
\end{equation}
where $t_\mathrm{c}$ and $\varphi_{\mathrm{c}}$ are the time and phase at coalescence, $\eta$ is the symmetric mass ratio, coefficients $\varphi_{i}$ and $\varphi_{i l}$ are constants determined by intrinsic parameters of binaries.

In order to capture potential deviation from GR, deformation coefficients 
\begin{equation} \label{tgr_para_gen}
    \{ \delta\varphi_0, \delta\varphi_2, \delta\varphi_3, \delta\varphi_4, \delta\varphi_{5 l}, \delta\varphi_6, \delta\varphi_{6 l}, \delta\varphi_7 \}
\end{equation}
are introduced through
\begin{equation} \label{generation_deformation}
    \varphi_{i} \rightarrow\varphi_{i}^{\mathrm{GR}}(1+\delta\varphi_{i}), \; \text{or}\ 
    \varphi_{i l} \rightarrow\varphi_{i l}^{\mathrm{GR}}(1+\delta\hat{\varphi}_{i l}).
\end{equation}
The deformation coefficient of 2.5 PN non-logarithmic term is not included due to its degeneration with coalescence phase.
Besides, following \citep{Datta2023} we have also not considered the 0.5 PN deformation coefficient $\delta\varphi_1$.
These deformation coefficients are treated as free parameters when performing parameter estimation. Potential departures from GR can be reflected by posteriors deviating from zero. 
In this work, intermediate and merger-ringdown deformation coffecients $\alpha_i$ and $\beta_i$ are not considered.

The model-agnostic phenomenological parameterized test of GW propagation utilizes a phenomenological modified GW dispersion relation \citep{Mirshekari2012} which takes the form of
\begin{equation} \label{dispersion}
    E^{2}=p^{2} + A_{\alpha} p^{\alpha},
\end{equation}
where $E$ and $p$ are the energy and momentum of GWs, $A_{\alpha}$ are phenomenological coefficients.
In this work, we follow tests performed by LVK \citep{Abbott2019b,Collaboration2020,Collaboration2021g} and consider cases of $\alpha$ running form 0 to 4 with cadence of 0.5 excluding the case of $\alpha=2$ where the speed of GW is independent with frequency. Thus the dephasing is a constant and completely degenerate with the arriving time of GW transients, and can only be constrained with the present of electromagnetic counterparts.

The adding power-law terms $A_\alpha p^\alpha$ in Eq. \ref{dispersion} make different frequency components of GWs propagate with different speed, and distort the observed GW waveforms.
Assuming the waveform in local wave zone is well-described by GR, only considering the dispersion effect during the propagation, the deformation in phase have been given as \citep{Mirshekari2012}
\begin{equation} \label{phase_deformation_prop}
    \delta\Phi_\alpha(f)=\mathrm{sign}(A_\alpha)
    \begin{cases}
    \frac{\pi (1+z)^{\alpha-1} D_\alpha}{(\alpha-1)}
    \lambda_\alpha^{\alpha-2} f^{\alpha-1},
    \quad&\alpha\neq1
    \\
    \frac{\pi D_\alpha}{\lambda_{A}}\ln (\pi \mathcal{M}f), &\alpha=1
    \end{cases},
\end{equation}
where $\lambda_A \equiv h |A_\alpha|^{1/(\alpha-2)}$, and $D_\alpha$ is defined as 
\begin{equation}
    D_\alpha = \frac{(1+z)^{1-\alpha}}{H_0}
    \int \frac{(1+z')^{\alpha-2} dz'}{\sqrt{\Omega_{\mathrm{M}} (1+z')^3 + \Omega_\Lambda}}.
\end{equation}
In the computation, we use $H_0= 67.66 \mathrm{km}/(\mathrm{Mpc} \cdot \mathrm{s})$, $\Omega_\mathrm{M}=0.3111$, and $\Omega_\Lambda=0.6889$ for the values of Hubble constant, matter and dark energy density parameters, which are taken from Planck 2018 results \citep{Aghanim2020}.
For the convenience to perform the multi-parameter test with PCA, we consider a parameterization slightly different with LVK.
The phase deformation Eq. \ref{phase_deformation_prop} can be rewritten as
\begin{equation} \label{deformation_propagation_2}
    \delta\Phi_\alpha(f)=
    \begin{cases}
    \delta\phi_\alpha
    \frac{(1+z)^{\alpha-1}}{(\alpha-1)\pi^{\alpha-2}}
    \left(\frac{10^7 M_\odot}{M}\right)^{\alpha-1}
    \left(\frac{D_\alpha}{1 \mathrm{Gpc}}\right)
    v^{3(\alpha-1)},\quad&\alpha\neq1
    \\
    \delta\phi_\alpha\frac{D_\alpha}{1\mathrm{Gpc}}\pi \ln(\pi \mathcal{M}f) , &\alpha=1
    \end{cases}.
\end{equation}
where
\begin{equation} 
    \delta\phi_\alpha = \frac{1{\rm Gpc} (10^7 M_\odot)^{1-\alpha}}{\lambda_A^{2-\alpha}}
\end{equation}
is the parameters varied in parameter estimation.

We employ the waveform model \texttt{IMRPhenomD} \citep{Husa2016, Khan2016} as the basis. The deformations of generation Eq. \ref{generation_deformation} are added into the PN expansion structure, and the deformation of propagation Eq. \ref{deformation_propagation_2} are added to the overall waveform.

\subsection{Bayesian parameter estimation} \label{subsec_bayes}

Bayesian parameter estimation is based on Bayes theorem
\begin{equation}
p(\boldsymbol{\theta}|\boldsymbol{d}, M) = \frac{\pi(\boldsymbol{\theta}|M) p(\boldsymbol{d}|\boldsymbol{\theta}, M)}{p(\boldsymbol{d}|M)},
\end{equation}
where $\boldsymbol{d}$ denotes observed data, $M$ and $\boldsymbol{\theta}$ represents the underlying model and parameters required to describe the model.
$p(\boldsymbol{\theta}|\boldsymbol{d}, M)$ is the posterior which  describes the probability distribution of the model parameters based on the observed data, and is the goal of parameter estimation. 
The posterior is a combination of two elements: the prior $p(\boldsymbol{\theta}|M)$ which implies our prior knowledge about the model parameters ahead of observation, and the likelihood $p(\boldsymbol{d}|\boldsymbol{\theta}, M)$ which represents the probability of a realization of time series observed by detectors given a particular set of model parameter values.
The evidence $p(\boldsymbol{d}|M)$ is a constant normalizing factor and can also be used in model comparison.

If the noise is stationary, Gaussian and uncorrelated, the likelihood can be written as \citep{Cutler1994}
\begin{equation}
p(\boldsymbol{d}|\boldsymbol{\theta}, M) \propto \exp \left[-\frac12 \sum_i \left\langle \boldsymbol{h}(\boldsymbol{\theta})- \boldsymbol{d}| \boldsymbol{h}(\boldsymbol{\theta})-\boldsymbol{d} \right\rangle \right],
\end{equation}
where $\boldsymbol{h}(\boldsymbol{\theta})$ is the detector responses for a waveform with a set of parameter values $\boldsymbol{\theta}$, $\boldsymbol{d}$ is the observed data, $i$ represent dependent measurements, here is different TDI channels which will be discussed in more detail below. 
The angle brackets denote the noise-weighted inner product defined as
\begin{equation}
\left\langle \boldsymbol{a}|\boldsymbol{b} \right\rangle = 4 \mathfrak{R} \int \frac{a(f)b^*(f)}{S_n(f)} \ {\rm d}f,
\end{equation}
with the noise power spectral density (PSD) $S_n(f)$. 
In this paper, we consider the noise model \texttt{SciRDv1} described in \citep{LISA2018}.

The response of space-borne detectors to GWs have many difference in various aspects comparing with ground-based detectors. 
For ground-based detectors, the wavelength of $\sim 500\mathrm{Hz}$ GWs is much longer than arm-length of detectors. The size of detectors can be neglected. For space-borne detectors, like LISA, the designed arm-length $2.5\times 10^6 \mathrm{km}$ corresponds to a frequency $f_* = 0.12 \mathrm{Hz}$ which is usually called the transfer frequency. If the frequency of GW is above the transfer function, in the journey of a laser photon from emission to reception, there may be multiple wavelengths of GW passing through the detector, which can result in the optical length variation canceled out by GW itself. 
To derive the response of a signal laser link, we need to perform integral of the metric perturbation along the geodesic of a photon \citep{Cornish2003,Krolak2004,Vallisneri2005}.
The response of a single link in frequency domain is given by 
\begin{equation} \label{single_links}
    \tilde{y}_{slr}(f)= i\pi f L \ 
    \mathrm{sinc} \left[\pi f L (1-\mathbf{k}\cdot\mathbf{n}_l)\right]
    \exp \left\{ -i \pi f [L + \mathbf{k}\cdot(\mathbf{p}_r + \mathbf{p}_s)] \right\}
    \mathbf{n}_l\cdot\mathbf{h}^{\rm TT}(f)\cdot\mathbf{n}_l.
\end{equation}
We follow the notation used in \citep{Marsat2021} where $l$ denote the link sent from node $s$ and received by node $r$, $\mathbf{p}_r$ and $\mathbf{p}_s$ are positions of the two nodes, unit vectors $\mathbf{k}$ and $\mathbf{n}_l$ point to the direction of the GW source and the direction of the laser link respectively.
It is an important difference of space-borne detectors comparing with ground-based detectors that the signal duration is longer during which the motion of detectors can not be neglected.
The vectors of detector constellation $\mathbf{n}_l$, $\mathbf{p}_s$ and $\mathbf{p}_r$ in Eq. \ref{single_links} are time-dependent.
For the signals like SMBHBs which will merge in the LISA sensitive band, the orbital evolution of GW source is much faster than the orbital evolution the detectors, the generalized stationary phase approximation can be used to obtain the single link response in frequency domain \citep{Marsat2018}, which directly substitutes the time in Eq. \ref{single_links} by the time-frequency correspondence
\begin{equation}
    t_f = -\frac{1}{2\pi} \frac{\mathrm{d} \Phi(f)}{\mathrm{d} f},
\end{equation}
where $\Phi$ is the phase of GW.
There is another difference that space-borne detectors are unequal-arm interferometers. The laser noise which is stronger than GW signals a few orders will experience different delays in the two arms, thus can not be canceled by itself at the photo detector backend.
TDI has to be used to construct observables by time-shifting and combining single link responses (Eq. \ref{single_links}) \citep{Tinto2020}.
The 1.5 generation TDI observable $X$ in frequency domain takes the form of \citep{Marsat2021}
\begin{equation}
    X = [y_{321} + z y_{123} - (y_{231} + z y_{132})] (1 - z^2),
\end{equation}
where $z=\exp [i 2\pi f L]$.
The other two observables $Y$, $Z$ can be obtained by cyclic permutation of indices.
The uncorrelated channels are given by combinations
\begin{equation}
    \begin{aligned}
        A &= \frac{1}{\sqrt{2}}(Z - X), \\
        E &= \frac{1}{\sqrt{6}}(X - 2Y + Z), \\
        T &= \frac{1}{\sqrt{3}}(X + Y + Z). \\
    \end{aligned}
\end{equation}
Although it is the unequal arm-length that that make it necessary to employ the TDI technique, in terms of noise feature of TDI observables, it is still proper to assume that the arm-length of each link is equal and constant \citep{Babak2021}. And with additional assumption that the noise of the same type have the same PSD, the noise of $A$, $E$, $T$ can be described by \citep{Babak2021}
\begin{equation}
    \begin{aligned}
        S_{A, E} &= 8\sin^2 (2 \pi f L) 
        {[2 + \cos (2 \pi f L)] S_{\rm OMS} + 
        [6 + \cos (4\pi f L) + 4\cos (2\pi f L)] S_{\rm acc}}, \\
        S_{T} &= 32 \sin^2 (2 \pi f L) \sin^2 (\pi f L)
        [S_{\rm OMS} + 4 \cos (\pi f L) S_{\rm acc}],
    \end{aligned}
\end{equation}
where $S_{\rm OMS}$ and $S_{\rm acc}$ denote the noise of optical metrology system and acceleration noise. We consider the noise model \texttt{SciRDv1} given in \citep{LISA2018} which reads
\begin{equation}
    \begin{aligned}
        \sqrt{S_{\rm OMS}}(f) &= 15\times10^{-12} \frac{2\pi f}c \sqrt{1+\left(\frac{2\times10^{-3}}f\right)^4} \left[\frac{1}{\sqrt{\mathrm{Hz}}}\right], \\
        \sqrt{S_{\rm acc}}(f) &= \frac{3\times10^{-15}}{2\pi f c} \sqrt{1+\left(\frac{0.4\times10^{-3}}f\right)^2}\sqrt{1+\left(\frac f{8\times10^{-3}}\right)^4} \left [ \frac 1 { \sqrt {\mathrm{Hz}} }\right].
    \end{aligned}
\end{equation}.

In practice, we use \texttt{lisabeta} \citep{Marsat2021} to compute detector responses and evaluate likelihood. We add the modification of Eq. \ref{generation_deformation} and Eq. \ref{deformation_propagation_2} into the GW phase and time-frequency relation based on the waveform model \texttt{IMRPhenomD} \citep{Husa2016, Khan2016}. To estimate the posterior, we employ the nested sampler \texttt{dynesty} \citep{Speagle2020} with MCMC evolution implemented in \texttt{bilby} \citep{Ashton2019}.
As mentioned in above, the response of space-borne detectors to GWs is more complex, and the longer signal duration leads to more frequency points required to be computed when evaluating the likelihood. The full Bayesian parameter estimation is extremely computational expensive for space-borne detectors. 
\texttt{lisabeta} \citep{Marsat2021} circumvent this problem by considering a noise-free likelihood and using interpolation on a sparse frequency grid.
While the likelihood without noise realization can not be used in analyses of real data, it can still capture the correlation between the parameters and multi-modality of posteriors.

\subsection{Principal component analysis} \label{subsec_pca}

Using Bayesian framework and stochastic sampling algorithm discussed in last subsection, we can obtain posterior samples whose densities can approximate to the posterior probability distributions of binary properties and deformation coefficients introduced in Section \ref{subsec_tgr}. However, varying all deformation coefficients simultaneously in parameter estimation can yields less informative posteriors due to correlation among them.
the PCA method can remedy this problem by transform deformation coefficients into new bases which are constructed by a linear combination of original deformation coefficients \citep{Datta2020,Pai2012,Saleem2021,Datta2022,Datta2023}. The new constructed parameters can be measured and constrained better by observed data, thus are more sensitive to potential deviation from GR. The procedure for PCA is briefly reviewed in following.

The covariance matrix for posterior samples of deformation coefficients obtained by Bayesian parameter estimation with marginalizing over GR parameters of binaries can be given by 
\begin{equation}
    C_{jk}=\left\langle \left(\delta\varphi_j - \langle\delta\varphi_j \rangle\right) \left(\delta\varphi_k-\langle\delta\varphi_k\rangle\right)\right\rangle.
\end{equation}
Here we use the deformation coefficients $\delta\varphi_i$ in GW generation test (Eq. \ref{tgr_para_gen}) for example, the procedure is same for $\delta\phi_\alpha$ in the test of GW propagation.
$\delta\varphi_j$ and $\delta\varphi_k$ denote posterior samples of different deformation coefficients and the angle brackets represent the expectation value of the random variable.
Diagonalizing the covariance matrix $C_{ij}$ of deformation coefficients, we can express this matrix by its eigenvalues and eigenvectors
\begin{equation}
    \mathbf{C}=\mathbf{U} \mathbf{S} \mathbf{U}^T,
\end{equation}
where $\mathbf{S}$ is a diagonal matrix with eigenvalues of $\mathbf{C}$ as its diagonal elements, $\mathbf{U}$ has eigenvectors of $\mathbf{C}$ as its columns.
The eigenvectors of $\mathbf{C}$ represent a new set of bases of deformation parameters, and original deformation parameters can be transformed into the new bases by 
\begin{equation}
    \delta\varphi^{\rm PCA}_{i\text{-th}}=\sum_{k}\alpha_{ik}\delta\varphi_{k},
\end{equation}
where $\alpha_{ik}$ are columns of $\mathbf{U}$, namely eigenvectors of covariance matrix $\mathbf{C}$, and $\delta\varphi_{k}$ are the original deformation coefficients. 
The eigenvectors of $\mathbf{C}$ define a set of new bases of deformation parameters. These new bases are different components of the posterior samples with different amounts of information, where the magnitude of each corresponding eigenvalue implies how principal the component is. Samples of dominant  components carry the most information of posteriors, thus have smaller variance and can be measured and constrained better. As we will see in next section, the dominant PCA parameters can be more sensitive to potential violations of GR.

\section{Results and discussions} \label{sec_res}
Following \citep{Datta2023}, we consider a SMBHB system with total mass of $7\times 10^5 M_{\odot}$ and mass ratio of 2. The aligned dimensionless spins of two components are $0.3$ and $0.2$ respectively. 
The detailed information of this system are presented in Table \ref{tab_SMBHB_info}.
As mention in Section \ref{subsec_bayes}, we add non-GR modification based on the waveform model \texttt{IMRPhenomD} in which the in-plane spins are not considered.
We consider two situations including tests with injections of GR and tests with injections of subtle deviations from GR. The detailed results are presented in following.

\begin{table}[]
    \centering
    \begin{tabular}{lcl}
        \toprule
        Parameters              & Injection values         & Priors \\
        \midrule
        total mass              & $7\times 10^5 M_{\odot}$ & Uniform[$6.5\times 10^5 M_{\odot}$, $7.5\times 10^6 M_{\odot}$] \\
        mass ratio              & 2.0                      & Uniform[1.5, 2.5]              \\
        aligned spin 1          & 0.3                      & Uniform[-1.0, 1.0]             \\
        aligned spin 2          & 0.2                      & Uniform[-1.0, 1.0]             \\
        luminosity distance     & 3 Gpc                    & Uniform[1 Gpc, 5 Gpc]          \\
        inclination             & 1.1 rad                  & Sine[0, $\pi$]                 \\
        polarization            & 1.7 rad                  & Uniform[0, $\pi$]              \\
        reference phase         & 1.2 rad                  & Uniform[0, 2$\pi$]             \\
        coalescence time        & 0.0                      & Uniform[-50, 50]               \\
        ecliptical longitude    & 0.8                      & Uniform[-$\pi$, $\pi$]         \\
        ecliptical latitude     & 0.3                      & Cosine[-$\pi$/2, $\pi$/2]      \\
        \midrule
        $\delta\varphi_0$     & 0.0008                     & Uniform[-0.01, 0.01]          \\
        $\delta\varphi_2$     & 0.1                        & Uniform[-1.5,  1.5]           \\
        $\delta\varphi_3$     & 0.6                        & Uniform[-7,    7]             \\
        $\delta\varphi_4$     & 5.0                        & Uniform[-60,   60]            \\
        $\delta\varphi_{5l}$  & 8.0                        & Uniform[-100,  100]           \\
        $\delta\varphi_6$     & 10.0                       & Uniform[-100,  100]           \\
        $\delta\varphi_{6l}$  & 15.0                       & Uniform[-100,  100]           \\
        $\delta\varphi_7$     & 16.0                       & Uniform[-180,  180]           \\
        \midrule
        $\delta\phi_{0.0}$    & 0.005,                    & Uniform[-0.06,0.06]            \\
        $\delta\phi_{0.5}$    & 0.03,                     & Uniform[-0.5, 0.5]             \\
        $\delta\phi_{1.0}$    & 0.2,                      & Uniform[-2,   2]               \\
        $\delta\phi_{1.5}$    & 0.3,                      & Uniform[-5,   5]               \\
        $\delta\phi_{2.5}$    & 1.2,                      & Uniform[-12,  12]              \\
        $\delta\phi_{3.0}$    & 1.8,                      & Uniform[-15,  15]              \\
        $\delta\phi_{3.5}$    & 1.6,                      & Uniform[-10,  10]              \\
        $\delta\phi_{4.0}$    & 0.2,                      & Uniform[-10,  10]              \\
        \bottomrule
    \end{tabular}
    \caption{{Information of injected signals and priors for parameter estimation. The first part shows the shared GR parameters, the second part is for the test of generation, and the third part is for the test of propagation. For the second and third parts, Only values used in tests with GR-violated singles are shown.}}
    \label{tab_SMBHB_info}
\end{table}

\subsection{Tests with injections of GR} \label{subsec_null}
\begin{figure}
    \centering
    \includegraphics[width=\columnwidth]{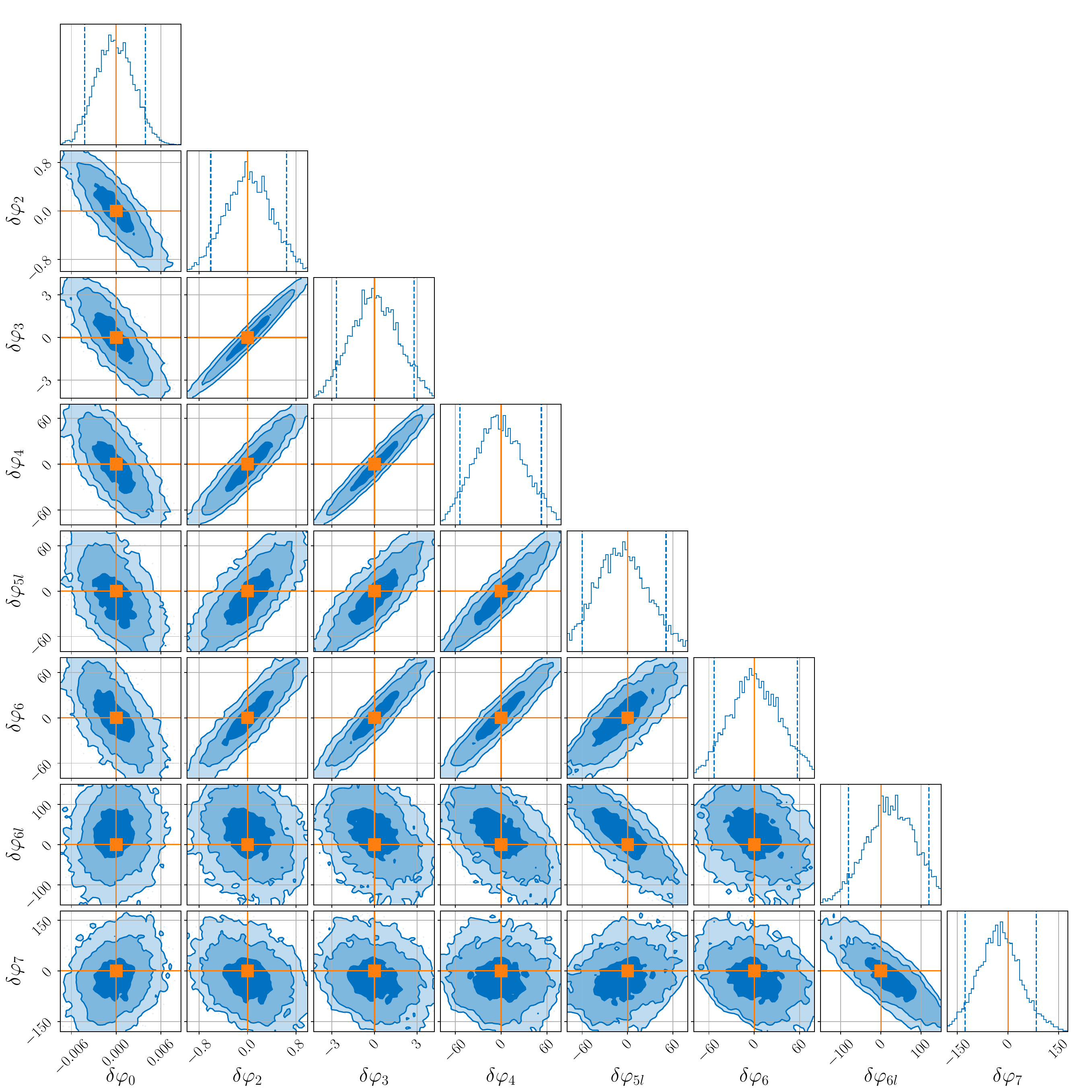}
    \caption{Posteriors of 8 PN deformation parameters after marginalizing over other GR parameters in the test of GW generation with the GR injection. The orange solid lines denote the injected GR values, and the dashed vertical lines denote $5\%$ and $95\%$ quantiles.}
    \label{fig_corner_pn}
\end{figure}

\begin{figure}
    \centering
    \includegraphics[width=\columnwidth]{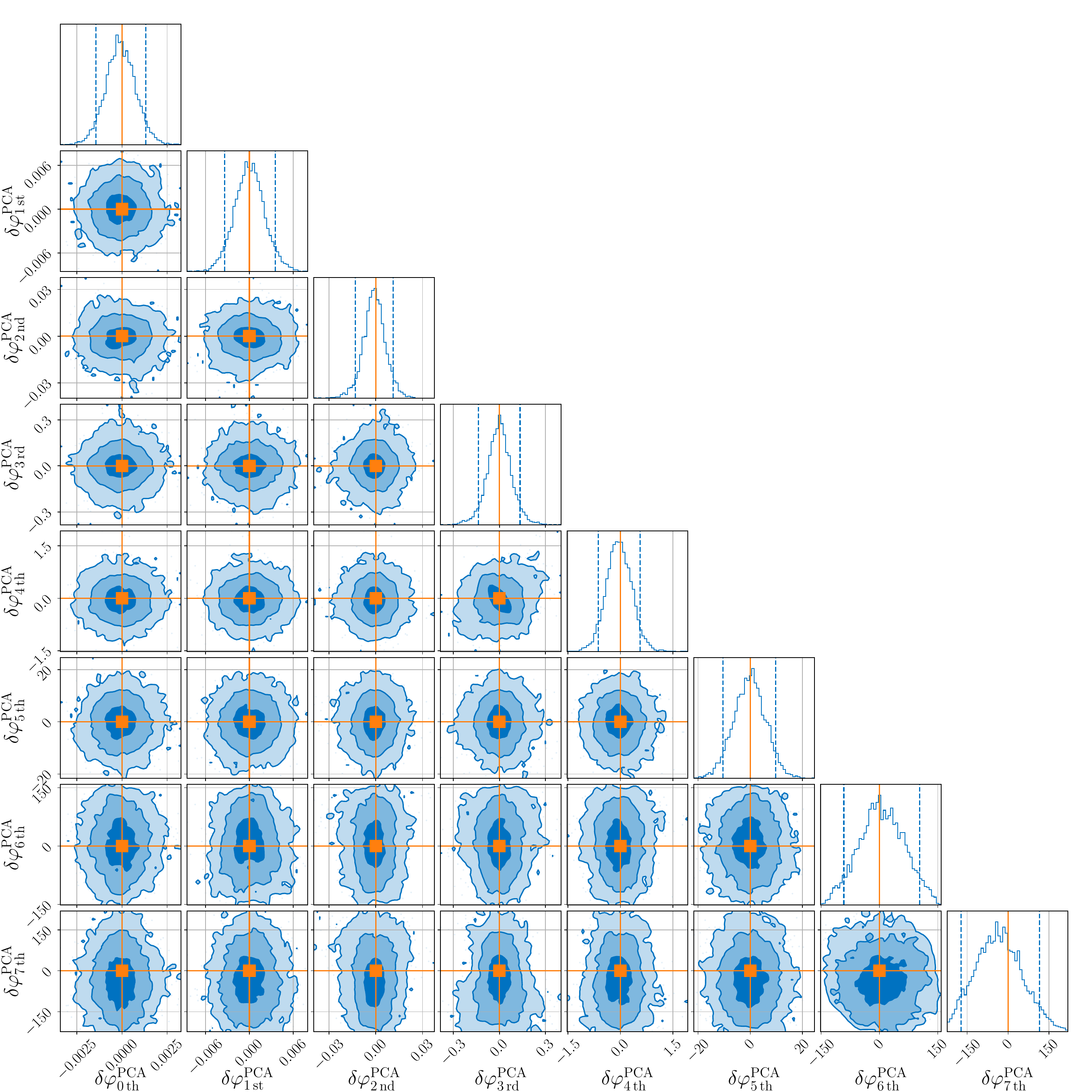}
    \caption{Probability distributions of PCA parameters which are obtained by transforming posterior samples of original PN deformation parameters shown in Fig. \ref{fig_corner_pn} into a set of new bases constructed by the PCA method. Comparing with Fig. \ref{fig_corner_pn}, it can be found the first few dominant PCA parameters can be better constrained and could be more sensitive to potential deviations from GR. The explicit values of $1\sigma$ bounds can be found in Tab. \ref{tab_null_pn}.}
    \label{fig_corner_pn_PCA}
\end{figure}

We first perform the test of GW generation with the GR injection. The GR signal is injected, while the waveform with deformation coefficients of Eq. \ref{tgr_para_gen} is used to recover properties of the binary. The 11 GR parameters and 8 deformation parameters are varied in parameter estimation. The posteriors of all deformation coefficients after marginalizing over other GR parameters are shown in Fig. \ref{fig_corner_pn}. 
Using the PCA method, the posterior samples can be transformed into a set of new bases. Results after this transformation are shown in Fig. \ref{fig_corner_pn_PCA}. 
For convenience of quantitative comparison, the standard deviations of posterior samples are summarized in Tab. \ref{tab_null_pn}.

As dicsussed in Sec. \ref{subsec_pca}, the PCA method transform original posterior samples into new bases. The amount of information carried by posteriors of each new bases is denoted by the magnitude of corresponding eigenvalues of the covariance matrix $\mathbf{C}$ of posterior samples.
The new bases of parameter space can be viewed as the components of posterior samples.
How informative the components are corresponds the size of error bars. The more dominant components can be measured and constrained better, and are more sensitive to penitential violations of GR. 
As can be seen by comparing Fig. \ref{fig_corner_pn} and \ref{fig_corner_pn_PCA}, the variance of the posterior of dominant PCA parameters shrinks significantly, even comparing the posterior of the most leading order PN deformation parameter.
{Quantitatively, the most dominant PCA parameter can be constrained to $\sim \mathcal{O}(10^{-4})$.}
This result is in the agreement with the previous expectations for multi-parameters tests of GW generation using Fisher matrix \citep{Datta2023} which reports the two most dominant PCA parameters can be bounded to $\sim \mathcal{O}(10^{-4})$.

\begin{figure}
    \centering
    \includegraphics[width=\columnwidth]{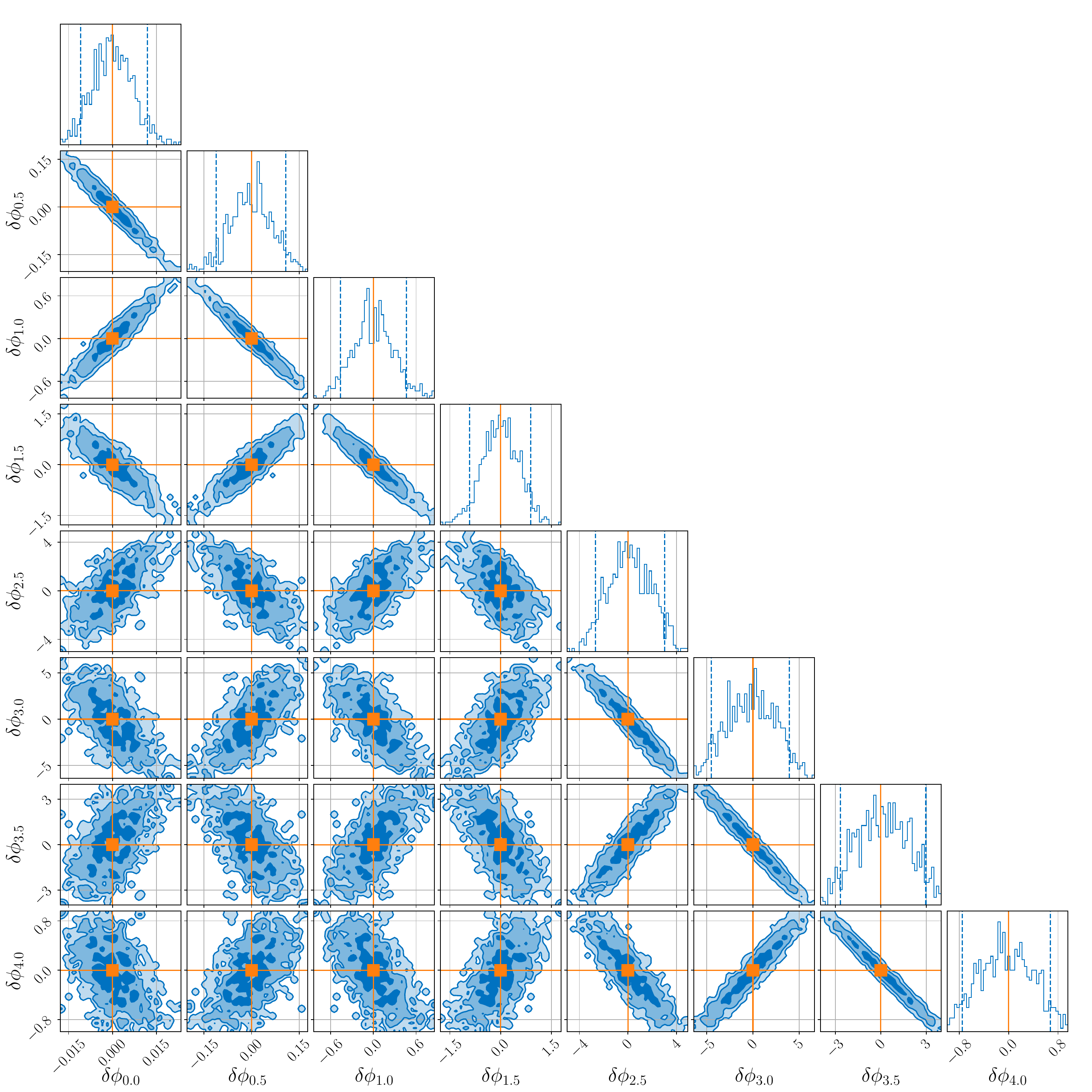}
    \caption{Posteriors of dispersion deformation parameters in the test of GW propagation with the GR injection. Same with previous cases, the orange solid lines and blue dashed lines denote injected GR values and quantiles of $5\%, 95\%$ repectively.}
    \label{fig_corner_dispersion}
\end{figure}

\begin{figure}
    \centering
    \includegraphics[width=\columnwidth]{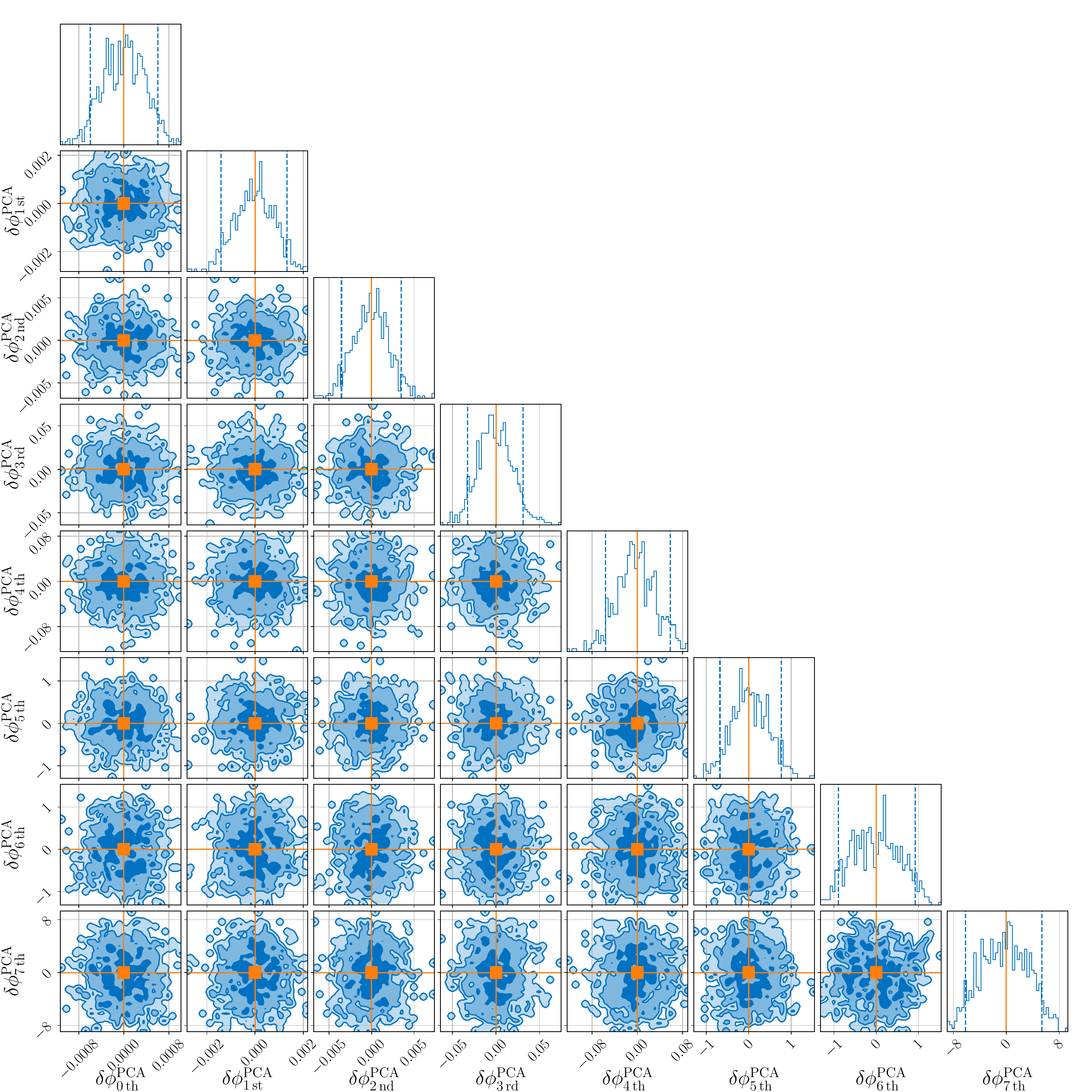}
    \caption{Distributions of posterior samples after PCA in the test of propagation with the GR injection.
    Through the procedure discussed in Sec. \ref{subsec_pca}, posterior samples of dispersion parameter can be transformed into a set of new bases where the information contained in posteriors is redistributed in different components. The first few dominant components include most information, thus have smaller deviations. These dominant components can be measured better by data, and are more sensitive to potential violations of GR.}
    \label{fig_corner_PCA_dispersion}
\end{figure}

Next, we focus on the test of GW propagation with the GR injection. The difference between the test of generation is that the deformation Eq. \ref{deformation_propagation_2} is added to the overall waveform rather the only PN expansion structure of inspiral.
The operation on posterior samples as discussed in Sec. \ref{subsec_pca} is the same with the test of generation.
The posteriors of original deformation parameters and the PCA parameters are shown in Fig. \ref{fig_corner_dispersion} and \ref{fig_corner_PCA_dispersion}. The standard deviations are collected in Tab. \ref{tab_null_dispersion} for quantitative comparison. 
Comparing with results of the test of generation, there is a difference that the constrains on original dispersion parameters are not monotonic with orders of frequency. This is due to the parameterization of Eq. \ref{deformation_propagation_2} we considered. Additional factors involving the $(\alpha-1)$ power term of source mass will change with the orders of frequency.
Observing the results of PCA parameters, the conclusion is the same with the test of generation.
The dominant PCA parameters can be measured better, and if there are penitential violations of GR, PCA parameters will be more sensitive to the departures.

From tests with injections of GR, we can find that the new parameters constructed by PCA method can be constrained better. Thus the PCA parameters are expected to be more sensitive to possible deviations from GR.
In order to elaborate the virtue of the PCA method more explicitly, we then consider the situation with injections of mild deviations from GR.

\begin{table}[]
    \centering
    \begin{tabular}{rlrl}
        \toprule
        \multicolumn{2}{c}{\bf PN deformation parameters} & \multicolumn{2}{c}{\bf PCA parameters} \\
        \midrule
        $\delta\varphi_{0}$  & 0.00246 & $\delta\varphi^{\rm PCA}_{0\,\rm th}$& 0.000841 \\
        $\delta\varphi_{2}$  & 0.372   & $\delta\varphi^{\rm PCA}_{1\,\rm st}$& 0.00207  \\
        $\delta\varphi_{3}$  & 1.62    & $\delta\varphi^{\rm PCA}_{2\,\rm nd}$& 0.00747  \\
        $\delta\varphi_{4}$  & 31.5    & $\delta\varphi^{\rm PCA}_{3\,\rm rd}$& 0.0834   \\
        $\delta\varphi_{5l}$ & 33.2    & $\delta\varphi^{\rm PCA}_{4\,\rm th}$& 0.369    \\
        $\delta\varphi_{6}$  & 32.7    & $\delta\varphi^{\rm PCA}_{5\,\rm th}$& 6.14     \\
        $\delta\varphi_{6l}$ & 59.8    & $\delta\varphi^{\rm PCA}_{6\,\rm th}$& 57.9     \\
        $\delta\varphi_{7}$  & 62.5    & $\delta\varphi^{\rm PCA}_{7\,\rm th}$& 85.1     \\
        \bottomrule
    \end{tabular}
    \caption{{Standard deviation $\sigma$ for posterior samples of PN deformation parameters and constructed PCA parameters in the test of GW generation with the GR injection. }
    }
    \label{tab_null_pn}
\end{table}

\begin{table}[]
    \centering
    \begin{tabular}{rlrl}
        \toprule
        \multicolumn{2}{c}{\bf dispersion parameters} & \multicolumn{2}{c}{\bf PCA parameters} \\
        \midrule
        $\delta\phi_{0.0}$  & 0.00687 & $\delta\phi^{\rm PCA}_{0\,\rm th}$ & 0.000370 \\
        $\delta\phi_{0.5}$  & 0.0661  & $\delta\phi^{\rm PCA}_{1\,\rm st}$ & 0.000828 \\
        $\delta\phi_{1.0}$  & 0.284   & $\delta\phi^{\rm PCA}_{2\,\rm nd}$ & 0.00210  \\
        $\delta\phi_{1.5}$  & 0.546   & $\delta\phi^{\rm PCA}_{3\,\rm rd}$ & 0.0203   \\
        $\delta\phi_{2.5}$  & 1.76    & $\delta\phi^{\rm PCA}_{4\,\rm th}$ & 0.0348   \\
        $\delta\phi_{3.0}$  & 2.59    & $\delta\phi^{\rm PCA}_{5\,\rm th}$ & 0.440    \\
        $\delta\phi_{3.5}$  & 1.71    & $\delta\phi^{\rm PCA}_{6\,\rm th}$ & 0.563    \\
        $\delta\phi_{4.0}$  & 0.429   & $\delta\phi^{\rm PCA}_{7\,\rm th}$ & 3.58     \\
        \bottomrule
    \end{tabular}
    \caption{{Similar to Tab. \ref{tab_null_pn}, standard deviation for posterior samples of GW dispersion parameters and their PCA parameters are shown here. 
    Due to the parameterization of Eq. \ref{deformation_propagation_2} considered here, where additional factors involving the $(\alpha-1)$ power term of source mass will change with the orders of frequency, the constrains on original dispersion parameters are not monotonic with orders of frequency like PN deformation parameters.}
    }
    \label{tab_null_dispersion}
\end{table}

\subsection{Detecting possible deviations from GR} \label{subsec_dev}
In the last subsection, it has been shown that dominant PCA parameters can be measured and constrained better, thus can be more sensitive to potential GR violations. In this subsection, we verify this conclusion by considering injections of subtle deviations from GR which may be difficult to detected by original deformation parameters.
But by using the PCA method, the degeneracy among deformation parameters can be broken in new constructed PCA parameters, thus the subtle departures in original parameters can leave significant indications in first few dominant PCA parameters.

\begin{figure}
    \centering
    \includegraphics[width=\columnwidth]{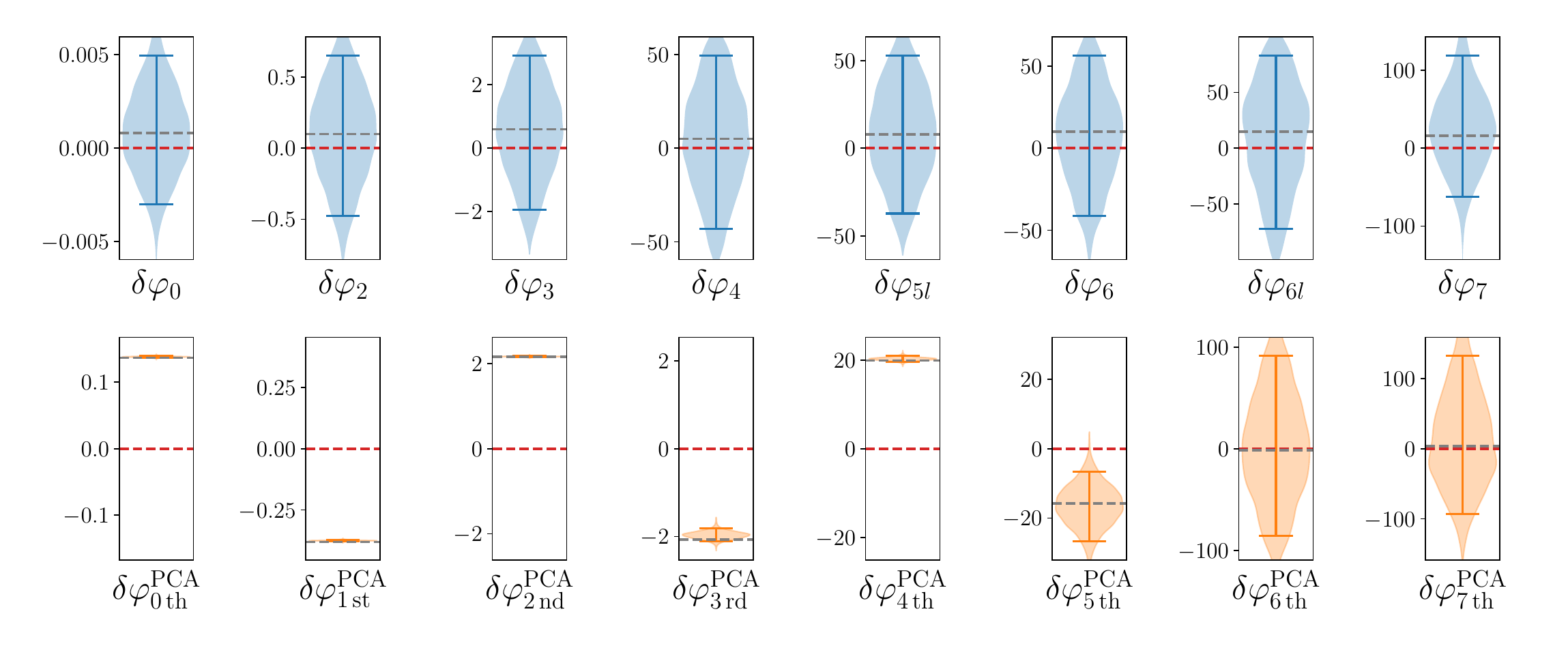}
    \caption{Violin plots for posteriors of PN deformation parameters in the test with injection of subtle GR violation. The red lines denote GR values and the gray lines denote injection values. The errors bars are determined by $5\%$ and $95\%$ quantiles of posterior samples. The original PN deformation parameters are shown by blue, while the PCA parameters are presented by orange. It can be clearly seen that small departures from GR can be identified by the PCA method in high significance.}
    \label{fig_violin_pn}
\end{figure}

\begin{figure}
    \centering
    \includegraphics[width=\columnwidth]{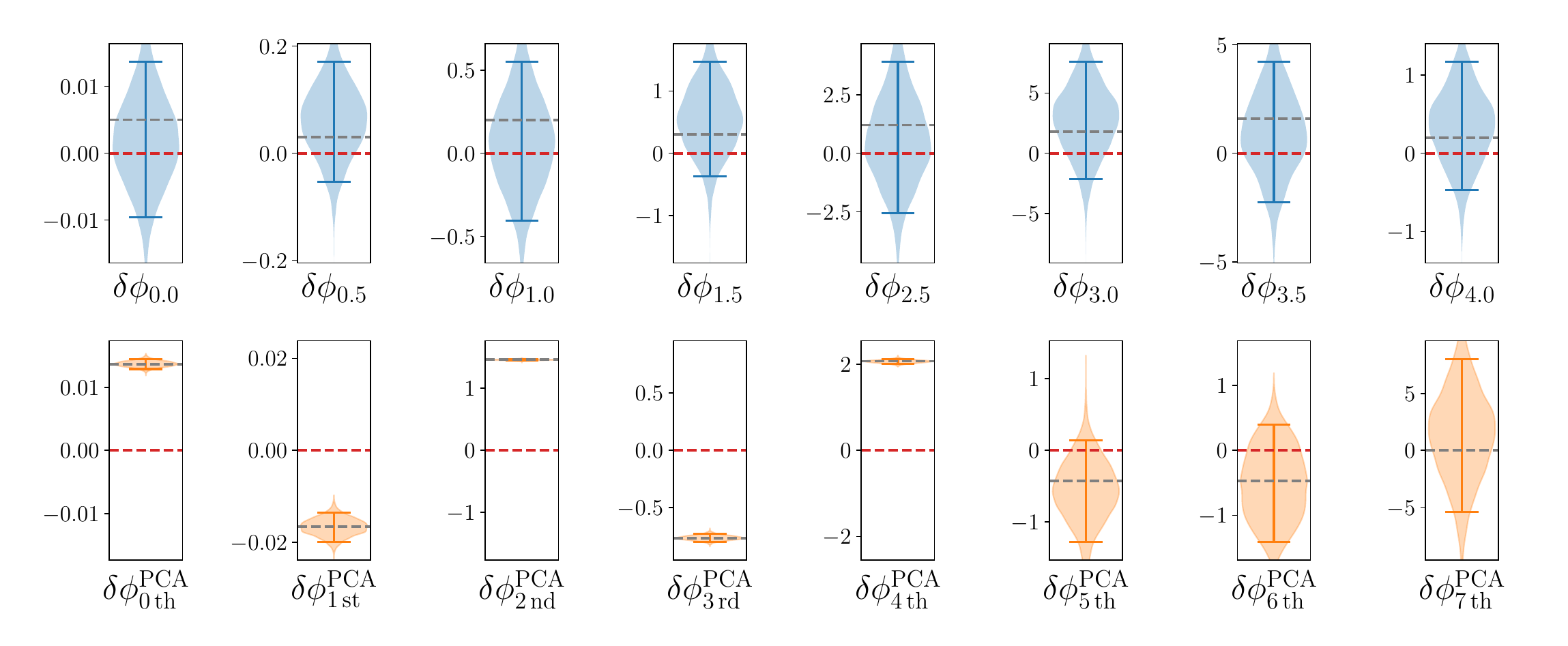}
    \caption{Similar to Fig. \ref{fig_violin_pn}, violin plots of the test for GW propagation with injection of subtle violations of GR. Small deviations in original dispersion parameters can yeild significant departures in first 5 dominant PCA parameters.}
    \label{fig_violin_dispersion}
\end{figure}

The methods for recovering source parameters and manipulating posterior samples are the same with tests with injections of GR. The difference is that injections of subtle GR violations are considered in this subsection.
As shown in Fig. \ref{fig_violin_pn}, the violin-plots with blue face represent posteriors of original PN deformation parameters where the gray dashed lines denote the injection values and red for GR values, the error bars are determined by $5\%$ and $95\%$ quantiles. The PCA parameters are shown by orange face. 
It can be clearly seen that the subtle GR violations which are difficult to be identified by original PN deformation parameters can lead significant departures in the first 5 dominant PCA parameters.
For convenience of quantitative comparison, the distance between GR values and posterior medians in the unit of the standard deviation of posterior samples are shown in Tab. \ref{tab_dev_pn}. The violations less than $1\sigma$ in original PN deformation parameters can lead significant departures in the first 5 dominant PCA parameters.
Similarly, the results for the test of propagation are presented in Fig. \ref{fig_violin_dispersion} and Tab. \ref{tab_dev_dispersion}. As expected, though the injections only have very slight GR violation, the first 5 dominant PCA parameters can significantly deviate from GR values, which demonstrates that the PCA parameters are more capable to capture potential GR violations.

\begin{table}[]
    \centering
    \begin{tabular}{rlrl}
        \toprule
        \multicolumn{2}{c}{\bf PN deformation parameters} & \multicolumn{2}{c}{\bf PCA parameters} \\
        \midrule
        $\delta\varphi_{0}$  & $0.348\sigma$ & $\delta\varphi^{\rm PCA}_{0\,\rm th}$&$    151\sigma$ \\
        $\delta\varphi_{2}$  & $0.307\sigma$ & $\delta\varphi^{\rm PCA}_{1\,\rm st}$&$   -172\sigma$ \\
        $\delta\varphi_{3}$  & $0.395\sigma$ & $\delta\varphi^{\rm PCA}_{2\,\rm nd}$&$    205\sigma$ \\
        $\delta\varphi_{4}$  & $0.156\sigma$ & $\delta\varphi^{\rm PCA}_{3\,\rm rd}$&$   -21.3\sigma$ \\
        $\delta\varphi_{5l}$ & $0.316\sigma$ & $\delta\varphi^{\rm PCA}_{4\,\rm th}$&$   48.7\sigma$ \\
        $\delta\varphi_{6}$  & $0.286\sigma$ & $\delta\varphi^{\rm PCA}_{5\,\rm th}$&$  -2.68\sigma$ \\
        $\delta\varphi_{6l}$ & $0.291\sigma$ & $\delta\varphi^{\rm PCA}_{6\,\rm th}$&$ 0.0232\sigma$ \\
        $\delta\varphi_{7}$  & $0.445\sigma$ & $\delta\varphi^{\rm PCA}_{7\,\rm th}$&$  0.132\sigma$ \\
        \bottomrule
    \end{tabular}
    \caption{The distance between GR values and posterior medians in the unit of standard deviation for the GW generation test with subtle GR violations are presented above, which can show the improvement of the capability to detect potential GR violations through the PCA method. 
    Since the degeneracy is broken in constructed PCA parameters, the subtle deviations in original PN deformation parameters can leave significant indication of departures in PCA parameters.}
    \label{tab_dev_pn}
\end{table}

\begin{table}[]
    \centering
    \begin{tabular}{rlrl}
        \toprule
        \multicolumn{2}{c}{\bf dispersion parameters} & \multicolumn{2}{c}{\bf PCA parameters} \\
        \midrule
        $\delta\phi_{0.0}$  & $0.246\sigma$  & $\delta\phi^{\rm PCA}_{0\,\rm th}$ & $  28.2\sigma$  \\
        $\delta\phi_{0.5}$  & $0.903\sigma$  & $\delta\phi^{\rm PCA}_{1\,\rm st}$ & $ -8.65\sigma$  \\
        $\delta\phi_{1.0}$  & $0.217\sigma$  & $\delta\phi^{\rm PCA}_{2\,\rm nd}$ & $   161\sigma$  \\
        $\delta\phi_{1.5}$  & $0.970\sigma$  & $\delta\phi^{\rm PCA}_{3\,\rm rd}$ & $ -35.9\sigma$  \\
        $\delta\phi_{2.5}$  & $0.258\sigma$  & $\delta\phi^{\rm PCA}_{4\,\rm th}$ & $  55.4\sigma$  \\
        $\delta\phi_{3.0}$  & $0.962\sigma$  & $\delta\phi^{\rm PCA}_{5\,\rm th}$ & $ -1.28\sigma$  \\
        $\delta\phi_{3.5}$  & $0.473\sigma$  & $\delta\phi^{\rm PCA}_{6\,\rm th}$ & $-0.920\sigma$  \\
        $\delta\phi_{4.0}$  & $0.708\sigma$  & $\delta\phi^{\rm PCA}_{7\,\rm th}$ & $ 0.361\sigma$  \\
        \bottomrule
    \end{tabular}
    \caption{Similar to Tab. \ref{tab_dev_pn}, the distance between GR values and posterior medians in the test of GW propagation are collected in this table for quantitative comparison.}
    \label{tab_dev_dispersion}
\end{table}

\section{Summary} \label{sec_summary}

GW observations provide new tools to explore the nature of universe. Current ground-based GW detectors have obtained fruitful accomplishments and been leading to a paradigm shift in researches of astrophysics, cosmology and gravity.
Space-borne detectors will open the window of low-frequency GW in the near future.
Testing gravitational theories is one of the most important scientific goals of space-borne detectors.
Previous works \citep{Datta2020,Pai2012,Saleem2021,Datta2022,Datta2023} have pointed out that using the PCA method can improve the capability of detecting potential violations of GR in multi-parameters tests.
In this work, we complement previous work \citep{Datta2023} by using Bayesian parameter estimation and extend this method to the test of GW propagation.

The phenomenological parameterized tests \citep{Mirshekari2012,Agathos2014} of GW generation and propagation have been routinely performed by LVK with current detected GW events. Due to correlations among deformation coefficients, varying all coefficients simultaneously can lead to less informative posteriors. Therefore, in the implementation of LVK \citep{Abbott2019b,Collaboration2020,Collaboration2021g}, only one deformation coefficient is allowed to vary at a time in parameter estimation.
However, it is proposed in \citep{Pai2012} that the PCA method can remedy this problem. A new set of bases are constructed in the PCA method, among which the information carried by posterior samples are redistributed. The dominant PCA parameters contain the most information of posteriors, thus have smaller variance, which means these parameters can be measured and constrained better, and be more sensitive to potential GR violations. The PCA method allow people to perform multi-parameter tests, i.e. varying all deformation parameters simultaneously, while still getting informative constrains on departures from GR through the dominant PCA parameters.

In this work, we consider multi-parameter tests with space-borne detector LISA, extend previous work \citep{Datta2023} by using Bayesian parameter estimation and apply the similar method to the test of GW propagation.
Following the previous work \citep{Datta2023}, we also consider a SMBHB system as GW source, and employ \texttt{lisabeta} to evaluate the likelihood which incorporates the motion of detector, the TDI combination, and the finite-size of arm-length. Nested sampler \texttt{dynesty} with sampling method implemented in \texttt{bilby} is used to estimate the posteriors of source properties. 
For the waveform model, we consider the phenomenological parameterized tests of GR generation and propagation based on \texttt{IMRPhenomD}. 
After the sampling process, the posterior samples of deformation coefficients are transformed to new bases given by the PCA method.
{The dominant new PCA parameters can be more sensitive to potential GR violation. }
We perform both tests with injections of GR and tests with subtle departures from GR.
From the obtained results, 
{in tests with injections of GR, we can find that the covariance of deformation parameters can be minimized by PCA method, thus the PCA parameters can be better measured.}
In the tests with subtle departures, the violations less than $1\sigma$ in original deformation parameters can yield significant departures in first 5 dominant PCA parameters.

It is worth noting that the phenomenological parameterized tests capture any violations of the waveform model used in parameter estimation. Not only anomalies beyond GR but also systematic errors of the waveform and unmodeled effects like eccentricity \citep{Garg2023,Saini2022,Bhat2022} or environmental effects \citep{Barausse2014}, etc. 
It is required further investigations to explore the influence of systematic errors and unmodeled effects on multi-parameter tests of GR.
Additionally, although the noise-free likelihood \citep{Marsat2021} considered in this work can capture the features of multi-modality and correlations among parameters, it can not be used to analysis real GW data. In future works, likelihood with noise realization and more realistic data analysis problems like signal overlapping, data gap, unstationary noise, etc. need to be considered.
{Moreover, One of drawbacks of the PCA method is that PCA parameters do not have any physical meanings and lose connections with specific alternative gravity theories unlike original PN deformation parameters which can be used to map constraints on specific theories through the parametrized post-Einsteinian (ppE) framework \cite{Yunes2009,Tahura2018}.
}

\section{Acknowledgements}
C.F. is supported by the National Key Research and Development Program of China Grant No. 2022YFC2204603 and by the starting grant of USTC.
R.N. is supported in part by the National Key Research and Development Program of China Grant No.2022YFC2807303.
W.Z. is supported by the National Key Research and Development Program of China Grant No.2021YFC2203102 and 2022YFC2200100, NSFC Grants No. 12273035 and 11903030, the Fundamental Research Funds for the Central Universities.
The numerical calculations in this paper have been done on the supercomputing system in the Supercomputing Center of University of Science and Technology of China.
Data analyses and results visualization in this work made use of \texttt{lisabeta} \citep{Marsat2021}, \texttt{Bilby} \citep{Ashton2019}, \texttt{Dynesty} \citep{Speagle2020}, \texttt{LALSuite} \citep{lalsuite}, \texttt{PESummary} \citep{Hoy2021a}, \texttt{NumPy} \citep{Harris2020, Walt2011}, and \texttt{matplotlib} \citep{Hunter2007}.

\appendix
\section{Corner plots for tests with injections of subtle GR-violated signals}
\begin{figure}
    \centering
    \includegraphics[width=\columnwidth]{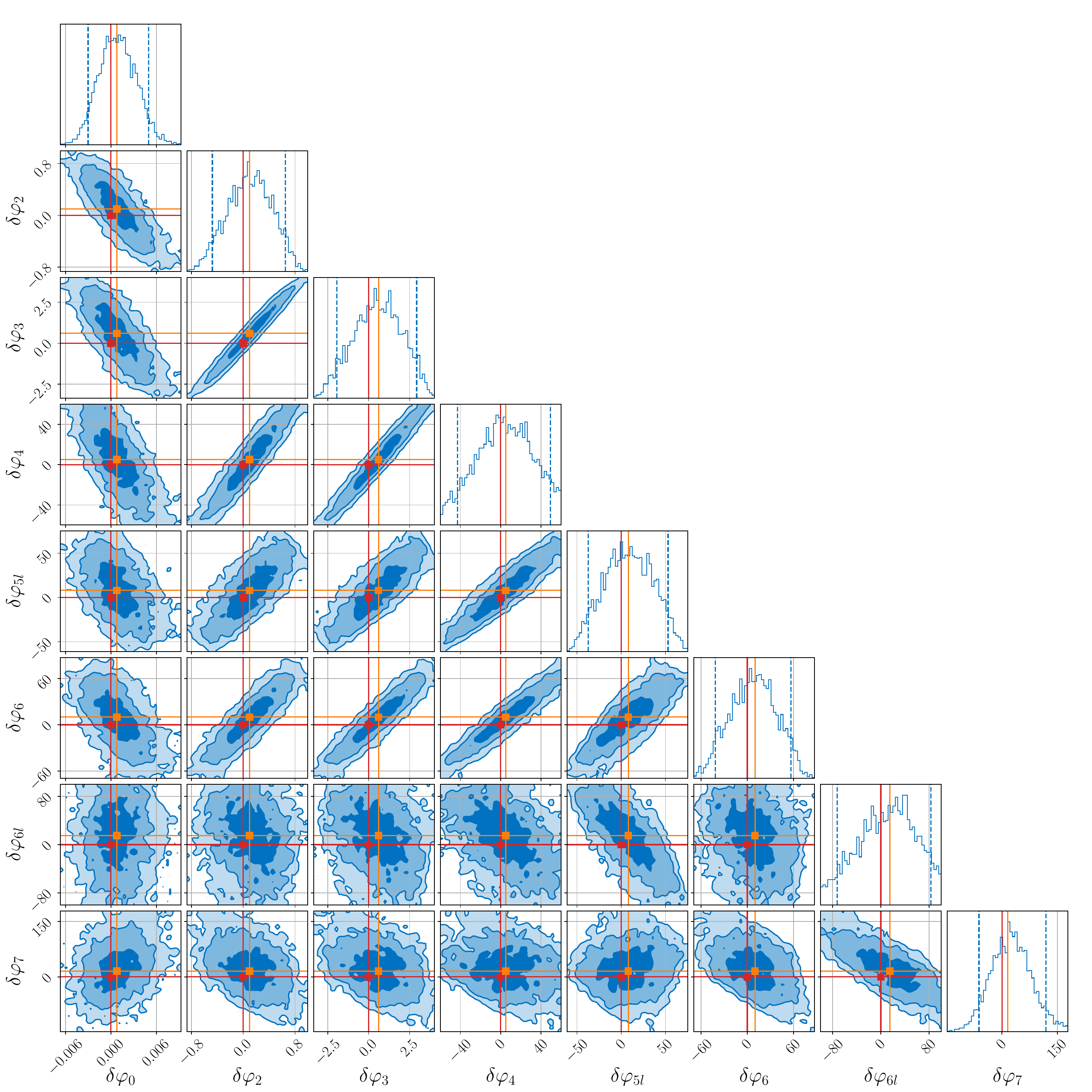}
    \caption{Posteriors of 8 PN deformation parameters in the test of generation with the injection of subtle GR-violated signal. Similar to Figure \ref{fig_corner_pn}, the orange lines denote the injection values, while the additional red lines denote GR values.}
    \label{fig_corner_pn_dev}
\end{figure}

\begin{figure}
    \centering
    \includegraphics[width=\columnwidth]{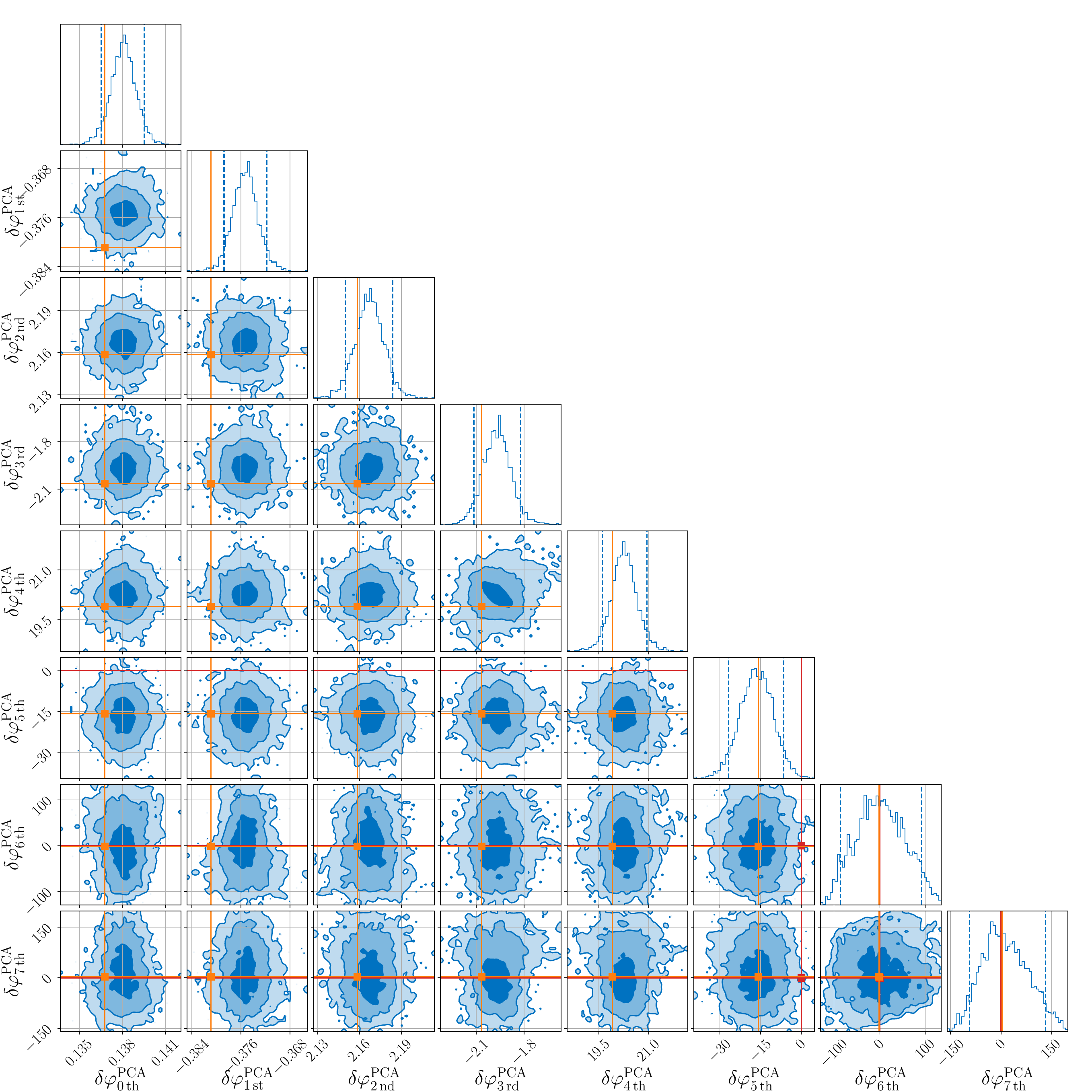}
    \caption{Results of applying PCA method to the posterior samples in the test of generation with injection of subtle GR-violated signal. Since the dominant PCA parameters are more sensitive to simulated deviations, the GR values are out of the plot ranges in the 5 most dominant components.}
    \label{fig_corner_pn_dev_pca}
\end{figure}

\begin{figure}
    \centering
    \includegraphics[width=\columnwidth]{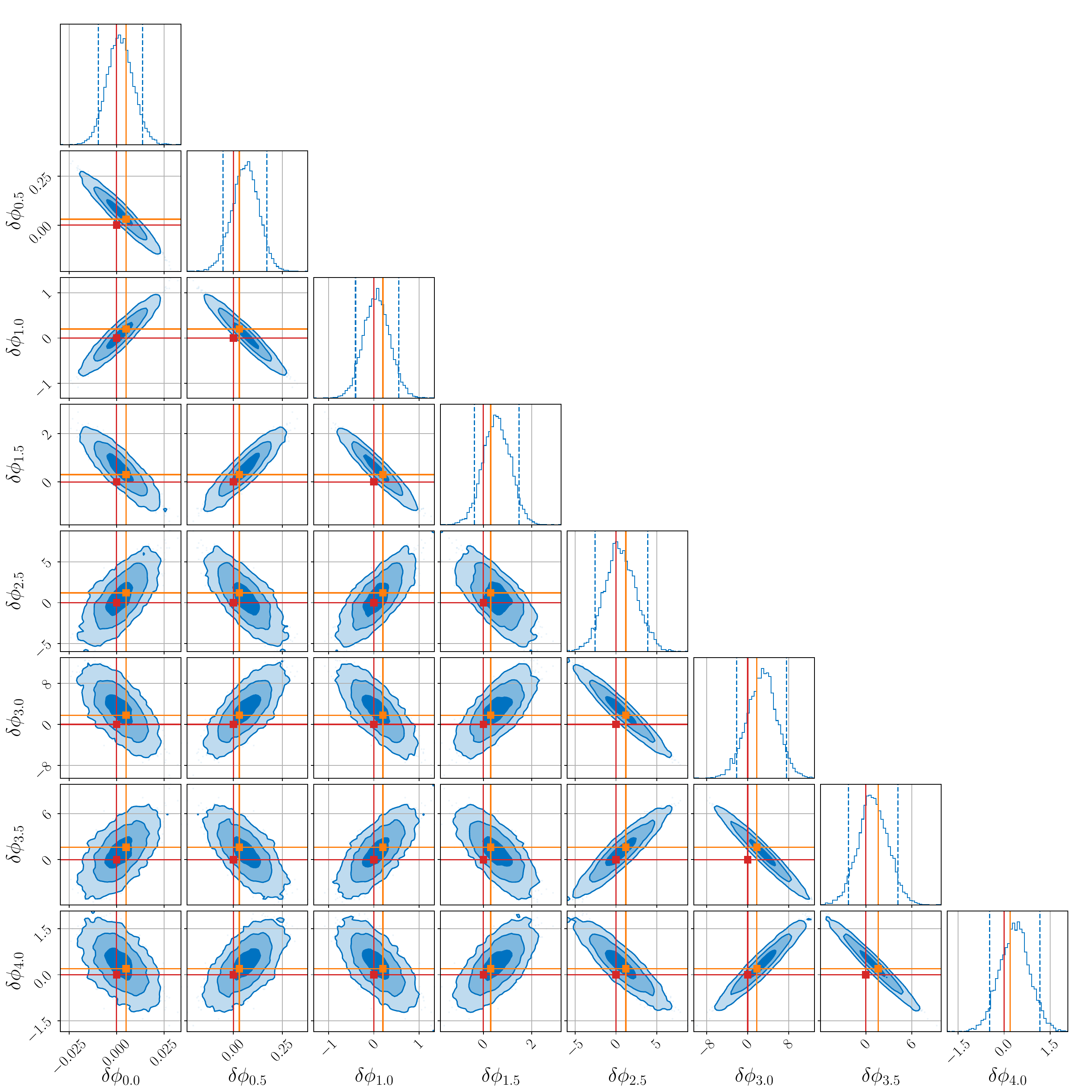}
    \caption{Posteriors of the test for GW propagation with the subtle GR-violated signals. The orange lines denote the injection values, and the red lines denote GR values.}
    \label{fig_corner_dispersion_dev}
\end{figure}

\begin{figure}
    \centering
    \includegraphics[width=\columnwidth]{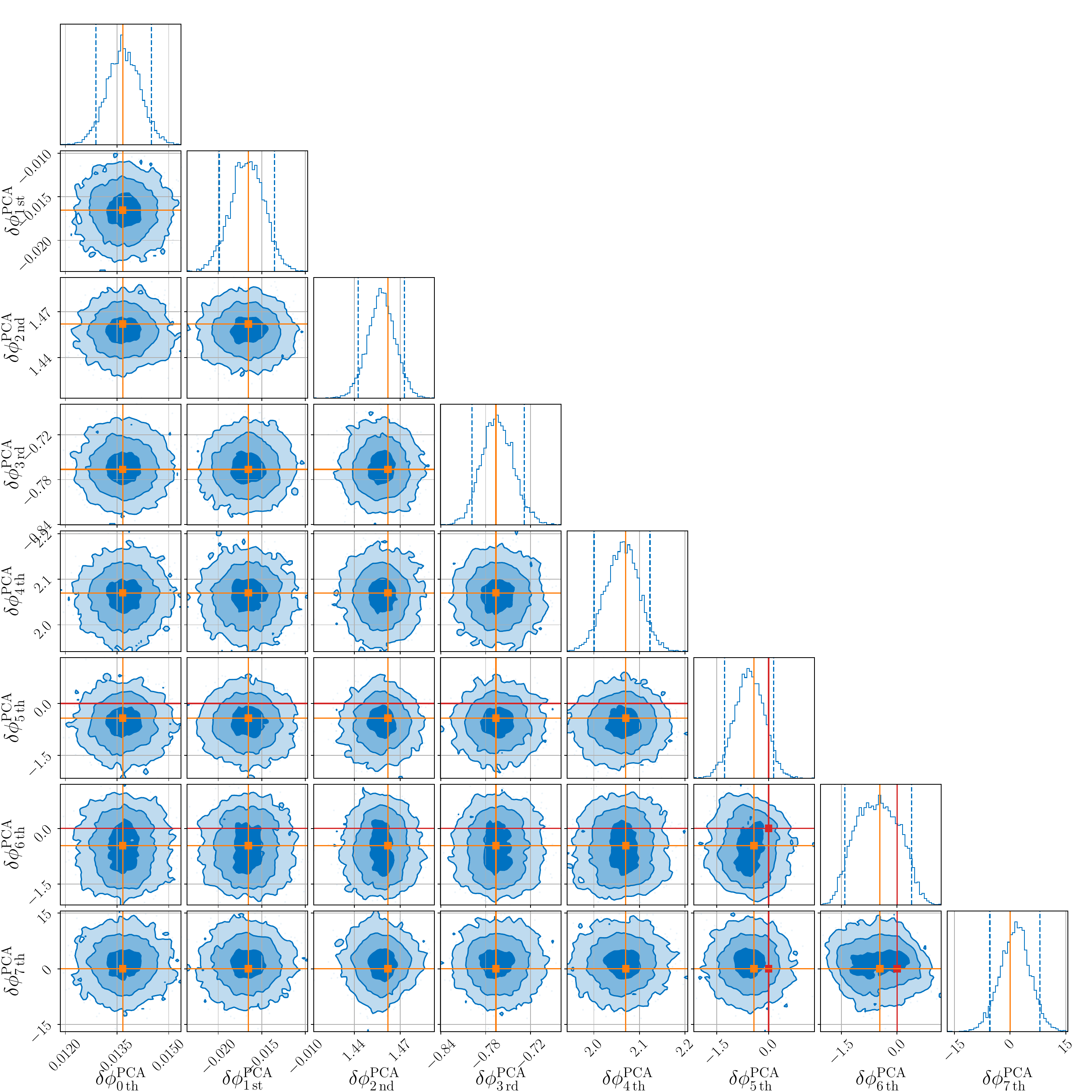}
    \caption{Results of applying PCA to the posterior shown in Figure \ref{fig_corner_dispersion_dev}. Similar to the case of generation. The GR values are out of plot ranges in the 5 most dominant PCA parameters.}
    \label{fig_corner_dispersion_dev_pca}
\end{figure}

{Similar to Figure \ref{fig_corner_pn}, \ref{fig_corner_pn_PCA}, \ref{fig_corner_dispersion}, and \ref{fig_corner_PCA_dispersion}, we also present the corner plots of posteriors for tests with injections of GR-violated signals as shown in Firgre \ref{fig_corner_pn_dev}, \ref{fig_corner_pn_dev_pca}, \ref{fig_corner_dispersion_dev}, and \ref{fig_corner_dispersion_dev_pca}.}


\newpage

\bibliographystyle{JHEP}

\bibliography{ref}



\end{document}